\documentclass[12pt]{article}
\usepackage{yfonts}
\usepackage{graphics}
\usepackage{bm}
\usepackage{graphicx}
\usepackage{amsmath}
\usepackage{amssymb}
\usepackage{amstext}
\usepackage{amscd}
\usepackage{amsfonts}
\usepackage{appendix}
\usepackage{color}
\DeclareMathAlphabet{\EuFrak}{U}{euf}{m}{n}
\DeclareMathAlphabet{\EuScript}{U}{eus}{m}{n}

\newcommand{\nd}{\noindent}

\newcommand{\be}{\begin{equation}}
\newcommand{\ee}{\end{equation}}
\newcommand{\ben}{\begin{eqnarray}}
\newcommand{\een}{\end{eqnarray}}

\title{{\bf MaxEnt, second variation, and  generalized statistics}}
\author{{A. Plastino$^1$ and M. C. Rocca$^1$} \\
\small{$^1$ La Plata National University and
  Argentina's National Research Council}\\
\small{(IFLP-CCT-CONICET)-C. C. 727, 1900 La Plata - Argentina}}

\date{\today}

\begin{document}

\maketitle

\begin{abstract}

\nd \small{ There are two kinds of Tsallis-probability
distributions: heavy tail ones and compact support distributions.
We show here, by appeal to functional analysis' tools, that for
lower bound Hamiltonians, the second variation's analysis of the
entropic functional guarantees that the heavy tail q-distribution
constitute a maximum of Tsallis' entropy. On the other hand, in
the compact support instance, a case by case analysis is necessary
in order to tackle the issue.}

\nd Keywords: MaxEnt, second variation, generalized statistics.

\end{abstract}

\newpage

\renewcommand{\theequation}{\arabic{section}.\arabic{equation}}

\section{Introduction}

\setcounter{equation}{0}

\nd During more than 25 years, an important topic in statistical
mechanics theory revolved around the notion of generalized
q-statistics, pioneered by Tsallis \cite{tsallis88}. It has been
amply demonstrated that, in many occasions,  the celebrated
Boltzmann-Gibbs logarithmic entropy does not yield a correct
description of the system under scrutiny \cite{tsallisbook}. Other
entropic forms, called q-entropies, produce a much better
performance \cite{tsallisbook}. One may cite a large number of
such instances. For example, non-ergodic systems exhibiting  a
complex dynamics \cite{tsallisbook}.

\nd The non-extensive statistical mechanics of Tsallis' has been
employed to fruitfully discuss phenomena in variegated fields. One
may mention, for instance, high-energy physics
\cite{[4]}-\cite{[44]}, spin-glasses \cite{[5]}, cold atoms in
optical lattices \cite{[6]}, trapped ions \cite{[7]}, anomalous
diffusion \cite{[8]}, \cite{[9]}, dusty plasmas \cite{[10]},
low-dimensional dissipative and conservative maps in  dynamical
systems \cite{[11]}, \cite{[12]}, \cite{[13]}, turbulent flows
\cite{[14]}, Levy flights \cite{wilk}, the QCD-based Nambu, Jona,
Lasinio  model of a many-body field theory \cite{15}, etc. Notions
related to q-statistical mechanics have been found useful not only
in physics but also in chemistry, biology, mathematics, economics,
and informatics \cite{[17]}, \cite{[18]}, \cite{[19]}.

\vskip 3mm
 \nd In this work we revisit the subject by appeal, in a classical MaxEnt phase-space framework,  to the second variation of
 functionals. We find that such analysis guarantees a maximum of
 the Tsallis' entropy only in the case of the heavy tail
 distributions. Our present treatment makes it advisable, on a more general
 MaxEnt framework, to always look at the second functional
 variation. We begin our discussion by remembering the concept of second variation.

\section{Second variation of a functional}

\setcounter{equation}{0}

\nd The essential concept that we need here is that of {\it
increment
 $h$ of a functional}. Note that the general theory of Variational
Calculus has been developed in a Banach Space (BS) \cite{shilov}.
Particularly important BS instantiations are, of course, Hilbert's
space and classical phase-space.
\setcounter{equation}{0}

\nd The MaxEnt approach in Banach space requires a first variation
that should vanish and a second one that ascertains the nature of
the pertinent extremum.  This second variation is not usually
encountered in MaxEnt practice, since one believes that the
entropy possesses a global maximum. This second functional
variation is the protagonist of the present endeavor. The approach
is  described in detail, for instance in the canonical book by
Shilov \cite{shilov} (for local minima). It is simply explained.

\nd One needs to evaluate the increment $h$ of a functional $F$ at
the point $y$ of the Banach space one is dealing with. One has
\begin{equation}
\label{eq2.1}
F(y+h)-F(y)=\delta^1F(y,h)+\frac {1} {2}
\delta^2F(y,h^2)+\varepsilon(h)
\end{equation}
where
\begin{equation}
\label{eq2.2} \lim_{h\rightarrow 0}\frac {||\varepsilon(h)||}
{||h||^2}=0.
\end{equation}
By  definition, $\delta^1F(y,h)$ is the first variation of $F$
 (if it is linear in $h$). $\delta^2F(y,h)$ is $F$'s second
 variation, quadratic in $h$. If $y$ is an extremum of $F$ then

\begin{equation}
\label{eq2.3}
\delta^1F(y,h)=0
\end{equation}
and it is a local minimum if
\begin{equation}
\label{eq2.4} \delta^2F(y,h)\geq C||h||^2\;\;\;\;\;C>0,
\end{equation}
or a local maximum if
\begin{equation}
\label{eq2.5}
\delta^2F(y,h)\leq C||h||^2\;\;\;\;\;C<0
\end{equation}
where  $C$ is a positive constant and  $||h||$ stands for the norm
of $h$. In phase-space, the object of our present concerns,
\begin{equation}
\label{eq2.6}
||h||^2=\int\limits_M h^2\;d\mu
\end{equation}
where  $M$ is the region of phase-space one is interested in and
$\mu$ the associated measure-volume for the concomitant space. We
start our consideration with reference to the orthodox instance.

\section{Motivation}

\setcounter{equation}{0}

\nd The following considerations should motivate the reader to
seriously consider the importance of elementary notions  of
functional analysis in q-statistics.

\subsection{q-Exponentials as linear functionals or distributions}

\nd A generalized function (or {\it distribution}) is a {\it
continuous functional} defined on a space of test-functions
\cite{gelfand}. A typical such test space is the so-called  ${\cal
K}$ space of Schwartz', of  infinitely differentiable functions
with compact support.

\nd One can prove \cite{gelfand} that $x_+^{\alpha}$,  defined by

\ben & x_+^{\alpha}=x^{\alpha}, \,\,for\,\, x>0, \cr &
x_+^{\alpha}=0, \,\,for\,\, x\leq 0. \een is a distribution
possessing single poles at integers $\alpha=-k$ with residues (at
the pole) \be R=\frac {(-1)^k} {(k-1)!}\delta^{(k-1)}(x),\ee with
$k=1,2,....,n,....$ \cite{gelfand}.

\nd A  function  is a particular instance of a distribution,
called {\it regular}  distribution. A singular distribution
 is that which cannot represented as a function. For example,  Dirac's delta is such a singular
 distribution.
 Tsallis'  q-exponentials $e_q$,  defined as
\be e_q(x)=[1+(q-1)x]_+^{\frac {1} {1-q}},\ee becomes clearly a
distribution  defined via

\ben & e_q(x)=[1+(q-1)x]_+^{\frac {1} {1-q}}= [1+(q-1)x]^{\frac
{1} {1-q}}\;\;{\rm if}\;\; 1+(q-1)x>0\cr &  e_q(x)=0\;\;\;\;\;{\rm
otherwise}.\een
 Computations involving $e_q$ should fruitfully appeal to distribution theory.

 \subsection{An instructive example for second variations
 \cite{shilov}}

\nd Generally, the minimal condition \be \delta^2 F(y,h)\geq
C||h||^2,\ee with  $C>0$ cannot be naively replaced by the weaker
restriction \be \delta^2 F(y,h)\geq.  0\ee

\nd Consider, for instance, minimizing  the  functional

\be
 F(y)=\int\limits_0^1y^2(x)[x-y(x)]\;dx.\ee
 It is easily seen that \be y(x)\equiv 0, \label{minim}\ee gives a functional extremum for $F(y)$, that is,
for $y$ given by (\ref{minim}) one has \be \delta^1F(y,h)=0.\ee
The second variation

\be
 \delta^2F(y,h)=\int\limits_0^1xh^2(x)\;dx, \ee is $ >0$ for any
function $h(x)\neq 0$. Thus, one may naively assume that
(\ref{minim}) yields a minimum for $F(y)$.

\nd To disprove such an assertion it is enough, given
$\epsilon>0$,  to consider as $y(x)$ any non-negative function
that is positive at $x=0$, does not exceed $\epsilon-x$ for
$x<\epsilon$, and vanishes for $x\geq \epsilon$. For example, let
$y(x)=\epsilon-x$ for  $x<\epsilon$ and $y(x)=0$ for $x\geq 0$.
Then,

\be F(y)=\int\limits_0^{\epsilon} (\epsilon-x)^2(2x-\epsilon)\;dx=
-\frac {\epsilon^4} {6} !\ee For a $y(x)\equiv 0$, $F=0$, but the
functional does not possess a minimum there.

\nd Similar considerations apply regarding local maxima apply if
one considers the restriction \be \delta^2 F(y,h)\leq C||h||^2,\ee
with  $C<0$, that cannot be replaced by the weaker condition \be
[\delta^2 F(y,h)\leq 0.\ee Choose, for instance, the  functional
\be F(y)=\int\limits_0^1y^2(x)[y(x)-x]\;dx, \ee and repeat the
above analysis.

\section{Applying second variation}

\setcounter{equation}{0}

\subsection{Boltzmann-Gibbs' Statistics}

\nd In the general case the prior information consists of $N$ mean
values corresponding to the observables $<R_i>,\,\,i=1, \ldots,N$:
However, the points we are about to make here emerge already at
the simplest level of just one observable, the Hamiltonian $H$
(canonical ensemble). We limit ourselves to this instance in this
work. Additionally, we assume that $H$ is lower bounded. The
MaxEnt variational functional becomes

\begin{equation}
\label{eq2.7} F_S(P)=-\int\limits_{M}P\ln(P)\;d\mu+
\alpha\left(\int\limits_{M}PH\;d\mu-<U>\right)+
\gamma\left(\int\limits_{M}P\;d\mu-1\right),
\end{equation}
with $P$ the probability density MaxEnt is designed to encounter.
$H$ is the hamiltonian whose mean value is called  $<U>$. Finally,
$\alpha$ and $\gamma$ will represent Lagrange multipliers. We
consider now $F_S$'s increment.

\[F_S(P+h)-F_S(P)=-\int\limits_{M}(P+h)\ln(P+h)\;d\mu+
\alpha\left[\int\limits_{M}(P+h)H\;d\mu-<U>\right]+\]
\[\gamma\left(\int\limits_{M}(P+h)\;d\mu-1\right)
+\int\limits_{M}P\ln(P)\;d\mu-
\alpha\left(\int\limits_{M}PH\;d\mu-<U>\right)-\]
\begin{equation}
\label{eq2.8} \gamma\left(\int\limits_{M}P\;d\mu-1\right).
\end{equation}
We can also write
\begin{equation}
\label{eq2.9} F_S(P+h)-F_S(P)=\int\limits_{M}
\left[\left(-1-\ln(P)+\alpha H +\gamma\right)h - \frac {h^2}
{2P}\right]\;d\mu+O(h^3).
\end{equation}
From  (\ref{eq2.9}) we find the first variation with its
associated Euler-Lagrange equation plus the second variation as
well. One has
\begin{equation}
\label{eq2.10}
-1-\ln(P)+\alpha H+\gamma=0
\end{equation}
\begin{equation}
\label{eq2.11}
-\int\limits_{M}\frac {h^2} {P}\;d\mu
\leq C||h||^2
\end{equation}
From (\ref{eq2.10}) one gathers that
\[P=\frac {e^{-\beta H}} {Z}\]
\begin{equation}
\label{eq2.12} Z=\int\limits_Me^{-\beta H}\;d\mu,
\end{equation}
with  $Z$ the system's partition function and $\beta$ proportional
to the inverse temperature. Eq. (\ref{eq2.11}) yields
\begin{equation}
\label{eq2.13} -\int\limits_{M}\frac {h^2} {P}\;d\mu=
-Z\int\limits_{M}h^2 e^{\beta H} \;d\mu  \leq
-Z\int\limits_{M}h^2\;d\mu= -Z||h||^2
\end{equation}
Notice that we can pass from the second to the third integral
because $e^{\beta H}$ is always greater than unity. {\it This is a
trivial point here, but not so when we consider Tsallis'
statistics below}. Looking at (\ref{eq2.13}) we see that one can
choose $C=-Z$. Remark that these are {\it classical}
considerations. Problems with (\ref{eq2.12}) at $T=0$ are thus not
surprising, on account of Thermodynamics' third law. As a bonus,
we discover here that the bound-constant $C$ is the partition
function itself.

\subsection{Tsallis' Statistics}

\nd Here there is no single way of computing mean values for the
theory \cite{tsallisbook}. Several options are available, that are
today considered equivalent for all practical purposes
\cite{tsallisbook,ferri}, because there is a ``dictionary" that
univocally relates two given probability densities $P_1\,-\,P_2$,
obtained using two different mean-values' choices \cite{ferri}.
 We consider in this work the three more important such choices and restrict ourselves to quadratic Hamiltonians.
 We insist in stating that the  q-exponential  is
 defined as ($x= \beta H \ge 0$)

 \ben \label{qexpo} & e_q(-x)= [1-(1-q)x]^{1/(1-q)},\, {\rm if}\,  (1-q)x \le 1\,;
 \cr & = 0 \,\,{\small {\rm otherwise\,(Tsallis' \,cutoff)}}, \een
that tends to the ordinary exponential as $q\rightarrow 1$. One
speaks of long tailed distributions for $[1-(1-q)x] \ge 0$ for all
$x>0$ and compact-support ones whenever the Tsallis cutoff becomes
operative for some $x-$values.
\subsection{Orthodox linear choice}

\nd One evaluates mean values in the customary fashion, linear in
$P$, i.e., \newline $<R>= \int_M R P \, d\mu$. The concomitant
Tsallis el functional is
\begin{equation}
\label{eq2.14} F_S(P)=-\int\limits_M\,
P^q\ln_q(P)\;d\mu+\alpha\left(
\int\limits_MPH\;d\mu-<U>\right)+\gamma
\left(\int\limits_MP\;d\mu-1\right),
\end{equation}
For the increment we have
\[F_S(P+h)-F_S(P)=-\int\limits_M(P+h)^q\ln_q(P+h)\;
d\mu+\alpha\left[
\int\limits_M(P+h)H\;d\mu-<U>\right]+\]
\[\gamma\left[\int\limits_M(P+h)\;d\mu-1\right]+
\int\limits_MP^q\ln_q(P)\;d\mu-\alpha\left(
\int\limits_MPH\;d\mu-<U>\right)-\]
\begin{equation}
\label{eq2.15}
\gamma\left(\int\limits_MP\;d\mu-1\right)
\end{equation}
Eq.  (\ref{eq2.15}) can be recast as (see Appendix B)
\[F_S(P+h)-F_S(P)=
\int\limits_M\left[\left(\frac {q} {1-q}\right)P^{q-1}
+\alpha H +\gamma\right]h\;d\mu-\]
\begin{equation}
\label{eq2.16} \int\limits_MqP^{q-2}\frac {h^2}
{2}\;d\mu+O(h^3),
\end{equation}
Eq. (\ref{eq2.16}) leads to the following equations:
\begin{equation}
\label{eq2.17} \left(\frac {q} {1-q}\right)P^{q-1} +\alpha H
+\gamma=0,
\end{equation}
\begin{equation}
\label{eq2.18} -\int\limits_MqP^{q-2}h^2\;d\mu\leq C||h||^2
\end{equation}
Eq. (\ref{eq2.17}) is the  Euler-Lagrange one while (\ref{eq2.18})
gives bounds originating from the second variation.  Thus,
(\ref{eq2.17}) entails (using the procedure given in \cite{tmp})
\begin{equation}
\label{eq2.19} \alpha=\beta q  Z^{1-q}
\end{equation}
\begin{equation}
\label{eq2.20} \gamma=\frac {q} {q-1}Z^{1-q}
\end{equation}
\begin{equation}
\label{eq2.21} P=\frac {[1+\beta(1-q)H]^{\frac {1} {q-1}}} {Z} =
e_{2-q}(-\beta H)/Z,
\end{equation}
\begin{equation}
\label{eq2.22} Z=\int\limits_M [1+\beta(1-q)H]^{\frac {1}
{q-1}}\;d\mu.
\end{equation}
 For Eq. (\ref{eq2.18}) we have,
\be \label{eq2.23}   W=-\int\limits_MqP^{q-2}h^2\;d\mu=
-\int\limits_MqZ^{2-q} [1+\beta(1-q)H]^{\frac {q-2} {q-1}}
h^2\;d\mu.  \ee
In order to obtain a bound, we need to find a constant $C$
(independent, in particular, of $H$). Thus, one needs to make sure
that the bracket \be \label{2} [1+\beta(1-q)H] \ge 0.\ee This
entails

\be \label{1} 0<q\le 1.\ee In such a case, the integral without
the bracket is smaller or equal than the integral with the bracket
and we have
 \be W   \le - qZ^{2-q} ||h^2||= -C ||h^2||. \label{5}\ee
 The present arguments, based on (\ref{2}),  {\bf guarantee} an entropic maximum only for long-tail (or heavy-tail)
Tsallis  distributions. For compact
 support distributions, the present arguments are inconclusive. The maximum may or may not exist. One should
  further investigate things on a case-by-case fashion. For instance, if $q>1$, the bracket $[1+\beta(1-q)H]$ might remain positive  for
low enough $\beta$. We will discuss this possibility in a
different Section below.



\subsection{Curado-Tsalllis mean values}
\nd A second alternative way of obtaining mean values has been
advanced in Ref. \cite{curado}, where the authors  define
\begin{equation}
\label{eq2.25} <U>=\int\limits_MP^qH\;d\mu,
\end{equation}
so that the MaxEnt Lagrangian $F_S$  becomes
\begin{equation}
\label{eq2.26} F_S(P)=-\int\limits_MP^q\ln_q(P)\;d\mu+\alpha\left(
\int\limits_MP^qH\;d\mu-<U>\right)+\gamma
\left(\int\limits_MP\;d\mu-1\right),
\end{equation}
and for the functional increment one writes (see Appendix B)
\[F_S(P+h)-F_S(P)=-\int\limits_M(P+h)^q\ln_q(P+h)\;
d\mu+\alpha\left[
\int\limits_M(P+h)^q H\;d\mu-<U>\right]+\]
\[\gamma\left[\int\limits_M(P+h)\;d\mu-1\right]+
\int\limits_MP^q\ln_q(P)\;d\mu-\alpha\left(
\int\limits_MP^q H\;d\mu-<U>\right)-\]
\begin{equation}
\label{eq2.27}
\gamma\left(\int\limits_MP\;d\mu-1\right)
\end{equation}
Expansion in order of up to $h^2$ is now demanded for A)
$(P+h)^q$, B) $\ln_q(P+h)$, and the product A) B). Keeping only
terms of order $h$ and $h^2$, one deduces from (\ref{eq2.27}) that
the linear term in $h$ yields
\begin{equation}
\label{eq2.28} -\frac {q} {q-1}[1-\alpha(q-1)H]P^{q-1}+\gamma=0,
\end{equation}
while de $h^2$-term generates
\begin{equation}
\label{eq2.29}
-\int\limits_Mq[1-\alpha(q-1)H]P^{q-2}h^2\;d\mu
\leq C||h||^2
\end{equation}
 Using (\ref{eq2.28}) produces here (with the procedure
 of \cite{tmp}):
\begin{equation}
\label{eq2.30}
\alpha=-\beta
\end{equation}
\begin{equation}
\label{eq2.31}
\gamma=-\frac {q} {q-1}Z^{1-q}
\end{equation}
\begin{equation}
\label{eq2.32} P=\frac {[1-\beta(1-q)H]^{\frac {1} {1-q}}} {Z}=
e_q(-\beta  H)/Z,
\end{equation}
\begin{equation}
\label{eq2.33} Z=\int\limits_M [1-\beta(1-q)H]^{\frac {1}
{1-q}}\;d\mu
\end{equation}
while (\ref{eq2.29}) leads to

\ben & W= -\int\limits_Mq[1-\alpha(q-1)H]P^{q-2}h^2\;d\mu=\cr & =
-\int\limits_MqZ^{2-q} [1 + (q-1)\beta H]^{\frac {q-2} {1-q}+1}h^2
=\cr & = -Z^{2-q}q \int\limits_M[1+\beta(q-1)H]^{\frac {1}
{q-1}}h^2\;d\mu. \een Again, in order to obtain a constant bound
we need the bracket in the last line above to be positive. This
entails, again, heavy-tail distributions, here implying

\be q \ge 1.  \label{3}\ee Thus,  \be \label{6}  C=qZ^{2-q}.\ee
The obvious demand $q>0$ is satisfied given (\ref{3}). As in the
preceding subsection, the present arguments {\bf guarantee} an
entropic maximum only for long-tail  Tsallis distributions. For
compact
 support ones, these arguments are inconclusive. The maximum may or may not exist. One should
  further investigate things on a case-by-case fashion. For instance, if $q<1$, the bracket $[1+\beta(q-1)H]$ might remain positive  for
low enough $\beta$. We will discuss this possibility in a future
Section below.

\subsection{Tsallis-Mendes-Plastino (TMP) mean values}

 \nd  Following Ref. \cite{tmp} we tackle the relationships
\begin{equation}
\label{eq2.35}
<U>=\frac {\int\limits_MP^qH\;d\mu} {\int\limits_MP^q\;d\mu}
\end{equation}
Our Lagrangian reads now
\[F_S(P)=-\int\limits_MP^q\ln_q(P)\;d\mu+\alpha\left(
\int\limits_MP^qH\;d\mu-\int\limits_MP^q\;d\mu <U>\right)+\]
\begin{equation}
\label{eq2.36} \gamma \left(\int\limits_MP\;d\mu-1\right).
\end{equation}
Thus (see Appendix B),
\[F_S(P+h)-F_S(P)=-\int\limits_M(P+h)^q\ln_q(P+h)\;d\mu+\]
\[\alpha\left[
\int\limits_M(P+h)^q H\;d\mu-
\int\limits_M(P+h)^q\;d\mu<U>\right]+
\gamma\left[\int\limits_M(P+h)\;d\mu-1\right]+\]
\[\int\limits_MP^q\ln_q(P)\;d\mu-\alpha\left(
\int\limits_MP^q H\;d\mu-
\int\limits_MP^q\;d\mu<U>\right)-\]
\begin{equation}
\label{eq2.37} \gamma\left(\int\limits_MP\;d\mu-1\right).
\end{equation}
This simplifies to
\[F_S(P+h)-F_S(P)=\]
\[\int\limits_M\left\{\left[-\frac {1} {q-1}+\alpha
(H-<U>)\right]qP^{q-1}+\gamma\right\}h\;d\mu+\]
\begin{equation}
\label{eq2.38} -\frac {1} {2}\int\limits_M[1-\alpha(q-1)
(H-<U>)]qP^{q-2}h^2\;d\mu+O(h^3).
\end{equation}
Now we gather that
\begin{equation}
\label{eq2.39}
\left[-\frac {1} {q-1}+\alpha
(H-<U>)\right]qP^{q-1}+\gamma=0
\end{equation}
\begin{equation}
\label{eq2.40} -\frac {1} {2}\int\limits_M[1-\alpha(q-1)
(H-<U>)]qP^{q-2}h^2\;d\mu\leq C||h||^2.
\end{equation}
From  (\ref{eq2.39}) one finds, with the usual procedure
(see \cite{tmp}):
\begin{equation}
\label{eq2.41} \alpha=-\frac {\beta} {1-\beta(1-q)<U>},
\end{equation}
\begin{equation}
\label{eq2.42} \gamma=\frac {qZ^{1-q}-[1-\beta(1-q)<U>]}
{(q-1)[1-\beta(1-q)<U>]},
\end{equation}
\begin{equation}
\label{eq2.43} P=\frac {[1-\beta(1-q)H]^{\frac {1} {1-q}}} {Z},
\end{equation}
\begin{equation}
\label{eq2.44} Z=\int\limits_M [1-\beta(1-q)H]^{\frac {1}
{1-q}}\;d\mu.
\end{equation}
A bit of algebra produces, from the above relations,

\ben & W=  -\int\limits_M[1-\alpha(q-1) (H-<U>)]qP^{q-2}h^2\;d\mu=
\cr & =  q \int\limits_M \,\,-kT\frac{1-  (1-q)  H }{kT + (q-1)
<U>} P^{q-2}h^2\;d\mu, \een and setting $P^{q-2}=$
$Z^{2-q}e_q(-\beta H )^{q-2},$

\be W =  -q Z^{2-q}{ \int\limits_M \,\, \frac{[1+(q-1)\beta H
]^{1/(q-1)}}{1+ \beta (q-1)<U>}} h^2d\mu.\ee One sees that, as in
the two preceding instances, the bracket $[1+(q-1)\beta H ]$ must
be positive (long-tails!) so as to find a constant bound. This
entails

\be \label{4} q\ge 1, \ee and one has

\be \label{eq2.45}     W \le -q Z^{2-q} \int\limits_M \,\,\frac
{1} {1+\beta(q-1)<U>}||h||^2 \leq C||h||^2, \ee that is \be
\label{7} C= q\frac {1} {1+\beta(q-1)<U>} Z^{2-q},\ee and  we
obtain the required bound for the entropy to become maximal.
Moreover, from (\ref{eq2.45}), we get, once again, the here
redundant condition $q>0$. As in the preceding two subsections,
here our  arguments  {\bf guarantee} an entropic maximum only for
long-tail  Tsallis distributions. For compact
 support distributions, the present arguments remain inconclusive. The maximum may or may not exist. One should
  further investigate things on a case-by-case fashion. The
  comment made below Eq. (\ref{6}) above is also pertinent here.

\section{Example: the Harmonic Oscillator (HO)}

\setcounter{equation}{0}

\nd Consider the simple Hamiltonian (in phase space) $H=P^2+Q^2$.
We will reconfirm the second variation functional restrictions
encountered above in the concomitant three q-statistics cases.

\subsection{Linear constraint $<U>$}

\nd We start by remembering (\ref{eq2.16}), that required $q\le
1$. One has
\[-\int\limits_MqP^{q-2}h^2\;d\mu=
-\int\limits_MqZ^{2-q} [1+\beta(1-q)(P^2+Q^2)]_+^{\frac {q-2}
{q-1}} h^2\;d\mu\leq\]
\begin{equation}
\label{eq1} -\int\limits_MqZ^{2-q}
[1+\beta(1-q)(P^2+Q^2)]_+^{\frac {1} {1-q}} h^2\;d\mu \leq
-qZ^{2-q} \int\limits_M h^2\;d\mu=-qZ^{2-q}||h||^2,
\end{equation}
since \be [1+\beta(1-q)(P^2+Q^2)]_+^{\frac {1} {1-q}}\geq 1.\ee
 We reobtain the restriction on heavy-tail Tsallis distributions.

\subsection{Curado-Tsallis non linear constraints} \nd We recall
(4.35) and (4.36). We had here the restriction $q\geq 1$ and deal
now with

\[-\int\limits_Mq[1-\alpha(q-1)(P^2+Q^2)]P^{q-2}h^2\;d\mu=
-\int\limits_MZ^{2-q}q[1+\beta(q-1)(P^2+Q^2)]_+^{\frac {1} {q-1}}
h^2\;d\mu\leq\]
\begin{equation}
\label{eq2} -\int\limits_MqZ^{2-q} h^2\;d\mu= -qZ^{2-q}||h||^2
\end{equation}
Since $q\geq 1$ we have $[1+\beta(q-1)(P^2+Q^2)]_+^{\frac {1}
{q-1}}\geq 1$. Heavy tails once again!

\subsection{TMP constraints }

Here we must go back to (4.46) and (4.49). The operative
restriction is  $q\geq 1$. Accordingly,
\[-\int\limits_M[1-\alpha(q-1)
(P^2+Q^2-<U>)]qP^{q-2}h^2\;d\mu=\]
\[-\frac {qZ^{2-q}} {1+\beta(q-1)<U>}
\int\limits_M[1+\beta(q-1)(P^2+Q^2)]_+^{\frac {1} {q-1}}
h^2d\mu\leq\]
\begin{equation}
\label{eq3} -\frac {qZ^{2-q}} {1+\beta(q-1)<U>}||h||^2 \leq
C||h||^2.
\end{equation}
We need again to appeal to heavy tail distributions.

\section{The compact support instance}

\setcounter{equation}{0}

We now consider the case of compact support probabilistic
distributions in the formulation of  Curado-Tsallis (similar
arguments can be made for the other two possibilities). In such a
case we need to  satisfy, for a maximum, the relation
\begin{equation}
\label{es1}
-Z^{2-q}q \int\limits_M[1+\beta(q-1)H]^{\frac {1}
{q-1}}h^2\;d\mu\leq C||h||^2
\end{equation}
A maximum is not guaranteed if  $0<q<1$. Consider now in more
detail such a q-interval. $H$ must be  bounded by above in all
phase space. By choosing $\beta$ sufficiently small, the
bracket  above does take a minimum positive value $\Delta$ and
then we have
\begin{equation}
\label{es2}
-Z^{2-q}q \Delta^{\frac {1} {1-q}}||h||^2\leq C||h||^2
\end{equation}
Thus, selecting
\begin{equation}
\label{es3}
-Z^{2-q}q \Delta^{\frac {1} {1-q}}=C
\end{equation}
we conclude that entropy does exhibit a maximum. Another way of
viewing this argument is to consider that
$H$ have a maximum value $R$. Consider $q<1$. More
specifically, $q= 0.5$. Then our critical bracket reads

\be  [1- 0.5  \beta R] > 0,   \ee entailing, for $\beta=1/kT$,

\be T> \frac{R}{2k}, \ee i.e., a {\it minimum} temperature in
order to guarantee the desired maximal condition that concerns us
here. One might wish to speculate that, for lower temperatures,
since no entropic maximum is possible, equilibrium might not be
reached.

\vskip 4mm

\nd  Let us, for instance, $H$ be given by
\begin{equation}
\label{es4}
H=(P^2+Q^2){\cal H}(P_0^2+Q_0^2-P^2+Q^2)
\end{equation}
where ${\cal H}$ is the Heaviside's step function. We need
$[1+\beta(q-1)H]$ to be positive.

\nd Selecting $q=\frac {1} {2}$ and $\beta=1/kT$ small enough we
have:
\begin{equation}
\label{es5} 0\leq P^2+Q^2\leq P_0^2+Q_0^2<\frac {2} {\beta},
\end{equation}
and thus

\be T > (P_0^2+Q_0^2)/2k ,    \ee

\begin{equation}
\label{es6}
\Delta=1-\frac {\beta} {2}(P_0^2+Q_0^2)
\end{equation}

\section{Conclusions}

\nd Our second variation bounds are given, for the three Tsallis'
cases, by the $C$-bounds (\ref{5}), (\ref{6}), and (\ref{7}),
respectively. Note that, since $Z$ vanishes at $T=0$, we can not
guarantee there a finite bound $C$, as required by the second
variation protocol.

\nd The three bounds yield exactly the same conclusion: entropic
maxima are guaranteed only for long tail Tsallis distributions.
 For compact
 support distributions, the present arguments are inconclusive. The maximum may or may not exist. One should
  further investigate things on a case-by-case fashion, as we have done, for particular instances, in Section 6.

\nd Summing up, the three Tsallis treatments, that were proved to
yield identical predictions for mean values in \cite{ferri}, still
give the same results concerning the requirements for entropic
maxima.

\nd  It is almost trivial to show that, for lower bounded
Hamiltonians, a quantum $n-$levels treatment yields identical
conclusions (see Appendix A).

\nd  Our present treatment makes it advisable, on a more general
 MaxEnt standpoint, to always look at the second functional
 variation.

 \newpage

\renewcommand{\thesection}{\Alph{section}}

\renewcommand{\theequation}{\Alph{section}.\arabic{equation}}

\setcounter{section}{1}

\section*{AppendixA}

\setcounter{equation}{0}

\nd We consider a quantum system with $n$ discrete levels of
positive energies $\epsilon_i$, probabilities $p_i$, and
increments $h_i$. The probability-vector $P$ and increment-vector
$h$ belong to $l^2$. Consider, for instance, the orthodox linear
choice for mean values, i.e., one evaluates mean values in the
customary fashion, linear in the probabilities. The concomitant
functional is
\begin{equation}
\label{eq2.146} F_S(P)=-\sum\limits_{i=1}^n
p_i^q\ln_q(p_i)+\alpha\left(
\sum\limits_{i=1}^np_i\epsilon_i-<U>\right)+\gamma
\left(\sum\limits_{i=1}^np_i\right),
\end{equation}
For the increment we have
\[F_S(P+h)-F_S(P)=-\sum\limits_{i=1}^n(p_i+h_i)^q\ln_q(p_i+h_i)+
\alpha\left[\sum\limits_{i=1}^n(p_i+h_i)\epsilon_i-<U>\right]+\]
\[\gamma\left[\sum\limits_{i=1}^n(p_i+h_i)-1\right]+
\sum\limits_{i=1}p_i^q\ln_q(p_i)-\alpha\left(
\sum\limits_{i=1}^np_i\epsilon_i-<U>\right)-\]
\begin{equation}
\label{eq2.156} \gamma\left(\sum\limits_{i=1}^np_i-1\right)
\end{equation}
Eq.  (\ref{eq2.156}) can be recast as
\[F_S(P+h)-F_S(P)=
\sum\limits_{i=1}^n\left[\left(\frac {q} {1-q}\right)p_i^{q-1}
+\alpha\epsilon_i +\gamma\right]h_i\]
\begin{equation}
\label{eq2.166} -\sum\limits_{i=1}^nqp_i^{q-2}\frac {h_i^2}
{2}+O(h^3),
\end{equation}
Eq. (\ref{eq2.166}) leads to the following equations:
\begin{equation}
\label{eq2.176} \left(\frac {q} {1-q}\right)p_i^{q-1}
+\alpha\epsilon_i +\gamma=0,
\end{equation}
\begin{equation}
\label{eq2.186} -\sum\limits_{i=1}^nqp_i^{q-2}h_i^2\leq C||h||^2
\end{equation}
Eq. (\ref{eq2.176}) is the  Euler-Lagrange one while
(\ref{eq2.186}) gives bounds originating from the second
variation. Thus, (\ref{eq2.176}) entails
\begin{equation}
\label{eq2.196} \alpha=\beta q  Z^{1-q}
\end{equation}
\begin{equation}
\label{eq2.206} \gamma=\frac {q} {q-1}Z^{1-q}
\end{equation}
\begin{equation}
\label{eq2.216} p_i=\frac {[1+\beta(1-q)\epsilon_i]^{\frac {1}
{q-1}}} {Z}
\end{equation}
\begin{equation}
\label{eq2.226} Z=\sum\limits_{i=1}^n
[1+\beta(1-q)\epsilon_i]^{\frac {1} {q-1}}.
\end{equation}

\nd For the bound given by (\ref{eq2.186}) we have

\be W= -\sum\limits_{i=1}^nqp_i^{q-2}h_i^2=
-\sum\limits_{i=1}^nqZ^{2-q} [1+\beta(1-q)\epsilon_i]^{\frac {q-2}
{q-1}} h_i^2.\ee To obtain a constant bound, independent of the
$\epsilon_i$, we must demand $ [1+\beta(1-q)\epsilon_i] \ge 0$
(long tail distribution!). Accordingly,
\begin{equation}
\label{eq2.236} W= -\sum\limits_{i=1}qZ^{2-q}
[1+\beta(1-q)\epsilon_i]^{\frac {1} {1-q}} h_i^2\leq -qZ^{2-q}
\sum\limits_{i=1}^n h_i^2=-qZ^{2-q}||h||^2
\end{equation}
From (\ref{eq2.236}) we see that $C=-qZ^{2-q}$ and $q>0$. As
$q\leq 1$ we have finally for $q$ the bounds $0<q\leq 1$.

\newpage

 \setcounter{section}{2}

\section*{AppendixB}

\setcounter{equation}{0}

Consider a functional of $P$ called  $F_S(P)$,   given by
\begin{equation}
\label{b1} F_S(P)=-\int\limits_M P^q\ln_q(P)\;d\mu= \frac {1}
{q-1}+\int\limits_M\frac {P^q} {1-q}\;d\mu.
\end{equation}
Thus, we can write
\begin{equation}
\label{b2} F_S(P+h)=-\int\limits_M (P+h)^q\ln_q(P+h)\;d\mu= \frac
{1} {q-1}+\int\limits_M\frac {(P+h)^q} {1-q}\;d\mu.
\end{equation}
Accordingly,
\begin{equation}
\label{b} F_S(P+h)= \frac {1} {q-1}+\int\limits_M \frac
{P^q+qP^{(q-1)}h+\frac {q(q-1)} {2}P^{q-2}h^2} {1-q}\;d\mu+O(h^3)
\end{equation}

\end{document}